\documentclass[structabstract]{aa}
\usepackage{graphicx}

\usepackage{natbib}
\usepackage{pdflscape}
\titlerunning{Orientation of HYMORS}
 \authorrunning{HYMORS}
\title{Orientation of the cores of hybrid morphology radio sources}

\author{M. Ceg\l{}owski, M.\,P. Gawro\'nski \& M. Kunert-Bajraszewska \\
         }
\institute{Toru\'n Centre for Astronomy, N. Copernicus University, Gagarina 11, 87-100 Toru\'n, Poland\\
        E-mail: \email{magda@astro.uni.torun.pl}}

\abstract
{} 
{
The FR\,I/FR\,II dichotomy is a much debated issue in the astrophysics of
extragalactic radio sources. Study of the properties of HYbrid
MOrphology Radio Sources (HYMORS) may bring  crucial information and lead
to a step forward in understanding the origin of FR\,I/FR\,II dichotomy. HYMORS are a rare
class of double-lobed radio sources where each of the two lobes clearly exhibits a different FR
morphology. This article describes follow-up
high resolution VLBA observations of the five  discovered by us HYMORS. The main aim of the observations 
was to
answer the questions of  whether the unusual radio morphology is connected to the orientation of objects 
towards the observer.
}   
{The milliarcsecond-scale structures are good probes of the galactic medium, the possible 
asymmetries, and orientation of the central engines of the sources. We obtained 
the high resolution radio maps of five hybrid radio morphology objects with the VLBA at C-band and L-band. 
}
{
The cores of all five sources have been detected at both radio bands. Two of them
revealed milliarcsecond core-jet structures, the next two objects showed hints of parsec-scale
jets, and the last one remained point-like at both frequencies.
}
{
We compared properties of observed milliarcsecond structures of hybrid
sources with the larger scale ones previously detected with 
the VLA. We find that on both scales the fluxes of their central components
are similar,  which may indicate the lack of additional emission in the
proximity of the nucleus. This suggests that jets present on the
$\sim$1-10\,kpc scale in those objects are FR\,II-like.  

When possible, the detected core-jet structures were used for estimating the core's spatial orientation.  
The result is that neither the FR\,I-like nor the FR\,II-like side is preferred, which may suggest
that no specific spatial orientation of HYMORS is required to explain their
radio morphology. Their estimated viewing angles indicate they are unbeamed objects. 
The 178/151\,MHz luminosity of observed HYMORS
exceed the traditional FR\,I/FR\,II break luminosity, indicating they have
radio powers similar to FR\,IIs.
}

\date{Received  /
Accepted }

\begin{document}
\maketitle
\section{Introduction }

According to \cite{fr}, there are two distinct morphological classes of radio galaxies, based on
their extended radio emission: Fanaroff-Riley type I and type
II sources (FR\,Is and FR\,IIs). FR\,IIs are quite homogeneous in morphology,
and their main characteristic is the prominent hotspots at the outer edges of
radio lobes. FR\,I sources show much more diversity \citep{parma02}. About half of them have
double-lobed morphology similar to FR\,IIs, and the rest is a mixing of
narrow-angle or wide-angle tailed sources (NATs and WATs). The characteristic feature of FR\,I sources
is a lack of hotspots at the outer parts of the radio structure.  
\cite{fr} also introduced the radio luminosity criterion $L_{178 MHz} \sim$
$10^{25.5}$ W~H$z^{-1}$ $sr^{-1}$ to distinguish objects between FR\,I class, which falls 
bellow this threshold, and the FR\,II class that is placed above.  
It has been also suggested that those classes undergo different cosmological evolutions \citep{wall}.    

Though the cause of morphological FR\,I/FR\,II dichotomy is still unclear, it is worth noticing that
there are three main hypotheses that focus on this phenomenon. Several authors claim that it 
is connected with transition of the intrinsically supersonic jet to transonic/subsonic flow 
that is being decelerated due to entrainment of thermal plasma in the innermost region of the 
host galaxy \citep[e.g][]{komi94,bow,bick,kia}. A second hypothesis links 
aberration in morphology with a more profound issue of the central engine \citep{mei,zib} or 
the jet itself \citep{rey, ghi}. Last but not least, the scheme attributes FR\,I/FR\,II dichotomy to 
the intergalactic medium, which together with jet power shapes the object morphology 
\citep{gopkri91,gopkri96}. Combining those two properties may explain how the hotspot became 
relatively subsonic to ambient medium, which leads to disruption of the collimated jet due 
to its weakened Mach disk. \cite{gopwii}(hereafter GKW00) show that there is a rare group of objects 
\cite[$<1$\,\% of extended FIRST objects;][]{gawron06} that indeed exhibits two different 
morphological classes of the jets, and they named them HYMORS ({\bf HY}brid {\bf MO}rphology 
{\bf R}adio {\bf S}ource). They point out that the HYMORS class may be essential for 
understanding the FR\,I/FR\,II dichotomy. 
GKW00 explains that the existence of those objects speaks in favor of a hypothesis 
in which the medium and jet power play a dominant role in shaping radio source. 


\begin{table*}
\setcounter{table}{0}
\begin{center}
\caption[]{Basic parameters of observed HYMORS.}
\resizebox{\textwidth}{!} {
\begin{tabular}{c c c c c c c c c c c c c}
\hline
\hline
Source name  &RA & Dec & {\it z
}& $ S_{1.4 {\rm GHz}}$&  $S_{4.9 {\rm GHz}}$&  $\alpha^{1.4 {\rm GHz}}_{4.9
{\rm GHz}} $& $ S^{c}_{1.4 {\rm GHz}}$&  $S^{c}_{4.9 {\rm GHz}}$&   $\alpha^{1.4
{\rm GHz}}_{4.9 {\rm GHz}} $& $L_{178 MHz}$ \\
 & h~m~s & $\degr$~$\arcmin$~$\arcsec$ & & mJy & mJy  &
& mJy & mJy &     & [W/Hz]\\
(1)& (2)& (3) & (4)& (5)&(6)&(7)&(8)&(9)&(10)&(11)\\
\hline
J1154+513 & 11\,53\,46.43 & +51\,17\,04.1 & $0.31^{*}$    &33.4 	&24.7 	& 0.24 & 33.4 & 27,4 & 0.16 &  27.03  \\
J1206+503 & 12\,06\,22.39 & +50\,17\,44.3 & -	 &4.5 	& 5.8 	& -0.20 & 4.5 & 5,8 & -0.21 &  - &\\
J1313+507 & 13\,13\,25.78 & +50\,42\,06.2 & $0.56^{*}$	 &7 	&6.1 	& 0.11  & 7.0 & 6.1 & 0.11  & $27.28^{+}$\\
J1315+516 & 13\,14\,38.12 & +51\,34\,13.4 & $0.63^{*}$	 &16 	& 13.1	&  0.16 & 16 & 13 & 0.16  & $27.10^{+}$\\
J1348+286 & 13\,47\,51.58 & +28\,36\,29.6 & 0.74 & 4.7  & - 	&  -    & -	& 44.7 & -  & $27.59^{+}$\\
\hline
\end{tabular}
}
\end{center}  

Description of the columns:\\
(1) source name;
(2) source right ascension (J2000) extracted from FIRST;
(3) source declination (J2000) extracted from FIRST;
(4) redshift, photometric redshift taken from SDSS  \cite[Sloan Digital Sky Survey,][]{york}, marked as (*);
(5) total flux density at 1.4\,GHz taken from FIRST;
(6) total flux density at 4.9\,GHz taken from GB6;
(7)  spectral index between
1.4 and 4.9\,GHz calculated using flux densities in columns (5) and (6)
($S\propto\nu^{-\alpha}$);
(8) core flux density measured based on 1.4\,GHz VLA observations
described in G06;
(9) core flux density measured based on 4.9\,GHz VLA observations            
described in G06;
(10) core spectral index between 1.4 and 4.9\,GHz calculated using flux densities in columns (8) and (9)
($S\propto\nu^{-\alpha}$);
(11)  logarithm of the  luminosity at 178 MHz , or at 151 MHz if stated differently (+), calculated using 
the analytical luminosity distance approximation \cite{lumi}.

\label{table1}
\end{table*}  


\cite{gawron06} (hereafter G06) nearly  doubled the number of known HYMORS sources. 
Results presented in G06 strongly support the findings of GKW00, namely 
that there are two different kinds of jets in HYMORS. Therefore, the existence 
of a FR-dichotomy as a whole is difficult to reconcile with the explanations that posit only 
fundamental differences in the central engine. It instead seems that in the 
case of HYMORS one-sided jet disruption caused by propagation into a large-scale over-density 
of cold gas is present and that it has a significant impact on FR-dichotomy. A number of \emph{Chandra} 
observations of HYMORS support this scenario \citep[see e.g.][]{miller06,miller09,massaro09}.  
  
Here we describe follow-up observations of the cores of the hybrid
sources discovered by G06. High resolution C-band and L-band observations
were made using VLBA in order to determine the orientation of their central 
engines and compare it with the large-scale radio structures of those sources visible 
on the VLA maps.
       
This paper is organized as follow: Section 2 describes data reduction and observations, 
Section 3 refers to information concerning individual sources,  Section 4 presents the results, 
Section 5 contains discussion, and finally Section 6 summarizes the main conclusion of this work.  

Throughout the paper, we assume the cosmology with
${\rm
H_0}$=71${\rm\,km\,s^{-1}\,Mpc^{-1}}$, $\Omega_{M}$=0.27,
$\Omega_{\Lambda}$=0.73.


\begin{figure*}[t]
\centering
\includegraphics[width=\textwidth]{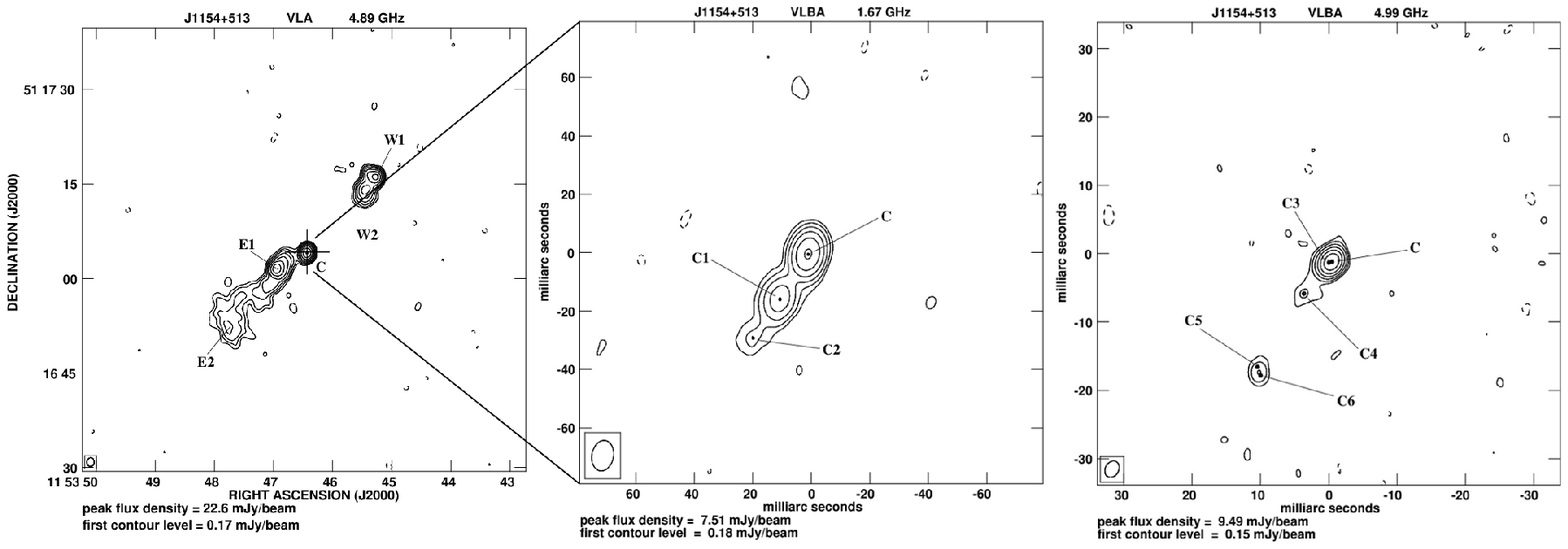} 
\includegraphics[width=\textwidth]{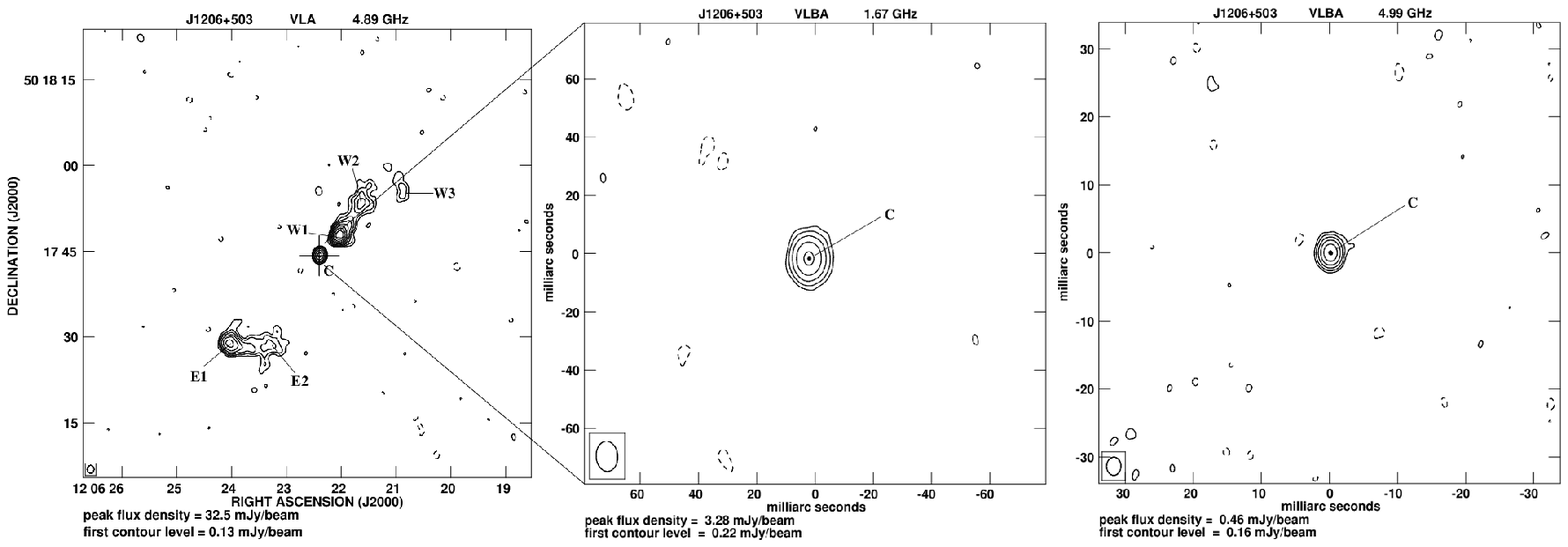}
\caption{
VLA C-band images reprinted from G06 (left) and VLBA L-band and C-band images
(middle and right) of J1154+513 \& J1206+503. Contours increase  
by a factor 2, and the first contour level corresponds to $\approx  
3\sigma$. A cross indicates the position of an optical object found using
the latest version of the SDSS.}   
\label{hybrids_1}
\end{figure*}


\section{Observations and data reduction}
In our previous work (G06), five new HYMORS were discovered using the VLA Faint
Images of the Radio Sky at Twenty-centimeters (FIRST) survey \citep{white97}. 
Additional 4.9 GHz VLA observations of the selected sources have
confirmed that indeed these are large-scale hybrids, with FR\,I and FR\,II
jet properties. The basic parameters of five target sources have been
gathered in Table~\ref{table1}.

The next step in our study was to look more close-by at the cores of those 	
uncommon sources and compare the orientation of their large-scale features
visible on the VLA maps with respect  to the identified cores. 

We made standard continuum observations of five HYMORS discovered by G06  
at the 1.7\,GHz and 5\,GHz using VLBA (26 \& 27 May 2007). Each target source, 
together with its associated phase reference source, was observed for $\sim$3\,hrs 
per frequency in phase-referencing mode of observations. Switching cycles of 8~min for
L-band and 5~min for C-band were used. 
The whole data reduction process was carried out using standard NRAO AIPS
procedures. IMAGR was used to produce the final total intensity images
(Fig.\,\ref{hybrids_1},\ref{hybrids_2},\ref{hybrids_3}). The flux densities of the main 
components of the target sources were then measured by fitting Gaussian models, using AIPS
task JMFIT. The properties of detected components are given in Table~2. If a given feature is identified
at both radio bands the spectral indexes are also calculated.

\section{Notes on individual sources}

\noindent {\bf J1154+513}~(Fig.\,\ref{hybrids_1})
Both 1.7\,GHz and 5\,GHz VLBA images show core-jet structures pointed towards FR\,I-like
large scale jet. Component C with a spectral index of $\alpha=-\,0.01$ visible
on both maps is a radio core. The components C1 - C6 are parts of the radio
jet. 
\vskip0.1cm

\noindent {\bf J1206+503}~(Fig.\,\ref{hybrids_1})
A single component is visible in both 1.7\,GHz and 5\,GHz VLBA images. Flat spectral index
$\alpha=\,0.0$ indicates it is a radio core.
\vskip0.1cm

\begin{figure*}[t]
\centering
\includegraphics[width=\textwidth]{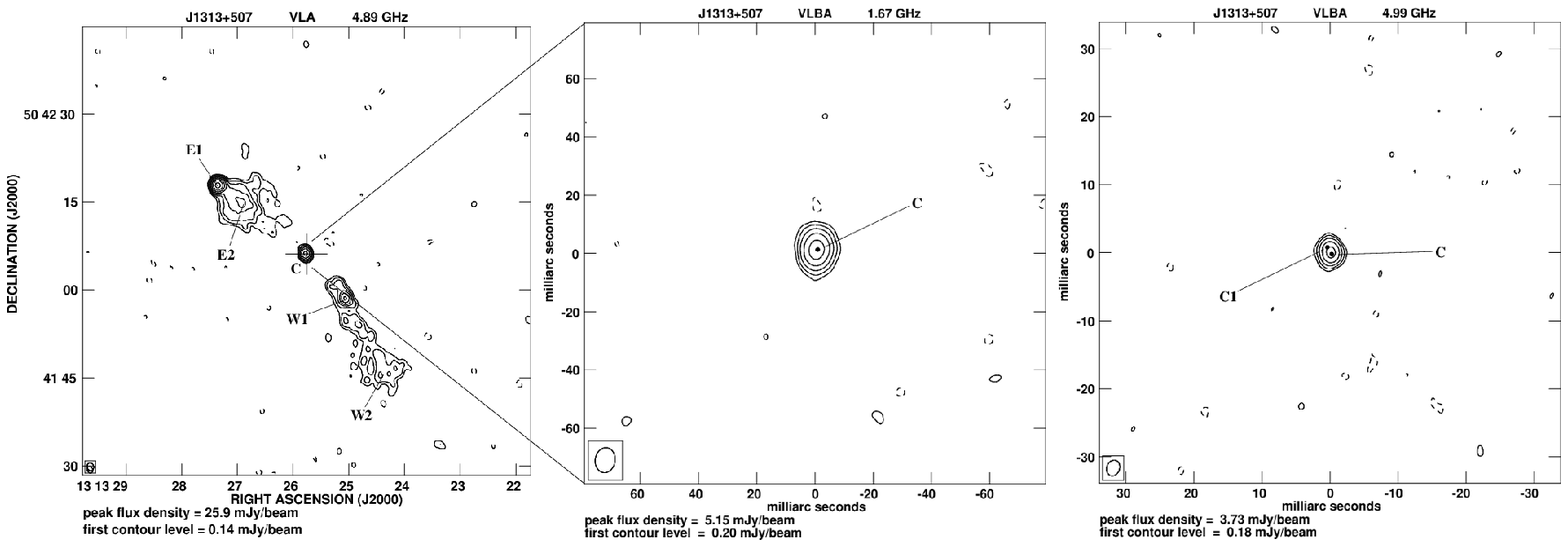} 
\includegraphics[width=\textwidth]{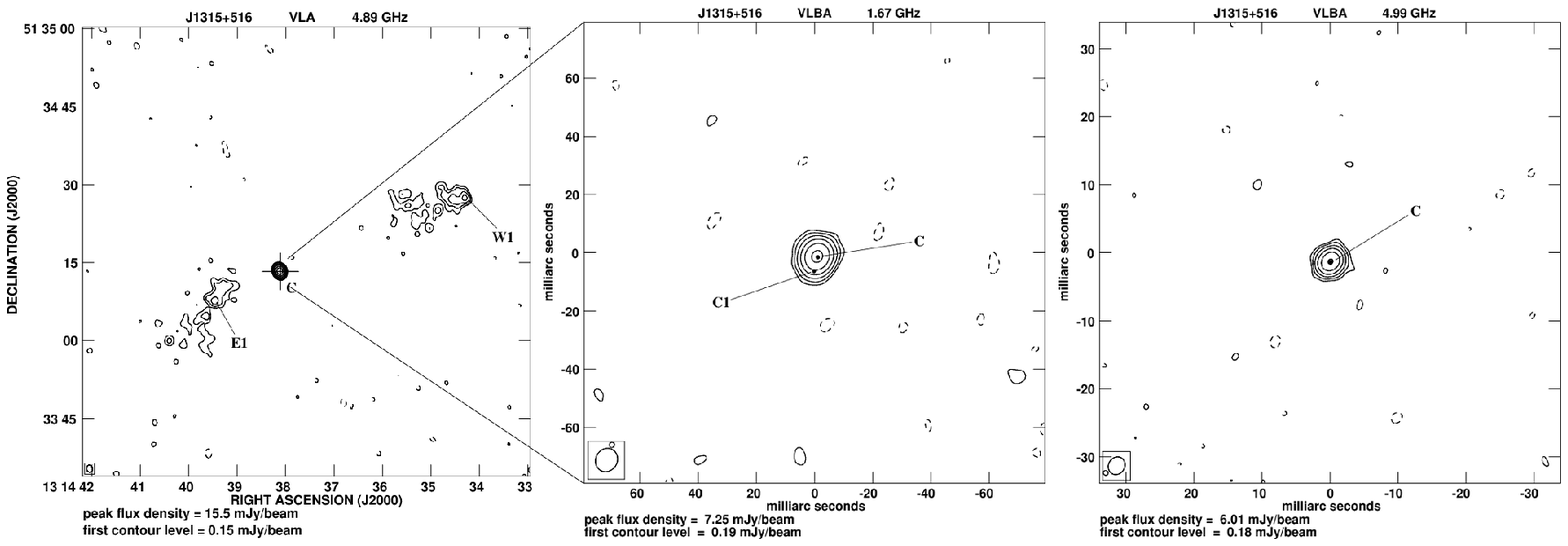}
\caption{
VLA C-band images reprinted from G06 (left) and VLBA L-band and C-band images
(middle and right) of J1313+507 \& J1315+516. Contours increase  
by a factor 2, and the first contour level corresponds to $\approx  
3\sigma$. A cross indicates the position of an optical object found using
the most actual version of the SDSS.}   
\label{hybrids_2}
\end{figure*}

\noindent {\bf J1313+507}~(Fig.\,\ref{hybrids_2})
The 5\,GHz image shows flat spectrum core C and an additional feature, probably a jet C1
directed towards FR\,II-like part of source. The core-jet structure
is unresolved in 1.7\,GHz observations. 
\vskip0.1cm

\noindent {\bf J1315+516}~(Fig.\,\ref{hybrids_2})
The 1.7\,GHz image revealed the probable presence of weak jet C1 directed towards the south;
however, the lack of C1 detection at C-band may indicate that this feature has a very
steep spectrum. The component C visible in both images is a radio core. 
\vskip0.1cm

\noindent {\bf J1348+286}~(Fig.\,\ref{hybrids_3})
Both VLBA images show the presence of a radio jet pointed towards an FR\,II-like part of the source. 
The structure of the jet is complicated, and in the 5\,GHz image, seven components could be distinguished. 
The flat spectrum core C has
inverted spectral index $\alpha=\,-0.29$. This source is classified as a quasar with 
a measured spectroscopic redshift of $z$=0.74 \citep{mun03}. 
\vskip0.1cm

\begin{table}[t]
\setcounter{table}{1}
\begin{center}
\caption[]{Flux densities of the principal components of the
sources measured based on the 1.7\,GHz and 5\,GHz VLBA observations
presented in this paper ($S\propto\nu^{-\alpha}$).}
\begin{tabular}{@{}c c c c c @{}}
\hline
\hline
Source & Compo- & \multicolumn{1}{c}{ $S_{1.7 {\rm GHz}}$} &
\multicolumn{1}{c}{ $S_{5 {\rm GHz}}$} &
\multicolumn{1}{r}{$\alpha^{1.7\,{\rm GHz}}_{5\,{\rm GHz}} $}  \\

name &nents  &\multicolumn{1}{c}{ mJy} &\multicolumn{1}{c}{ mJy}\\

(1)& (2)&  \multicolumn{1}{c}{(3)}&\multicolumn{1}{c}{(4)}
&\multicolumn{1}{c}{(5)}\\

\hline
J1154+513 & C  &  7.5 $\pm$ 0.1 &  7.6 $ \pm $  0.1 & -0.01\\
          & C1 &  2.9 $\pm$ 0.1 &  $-$              & $-$  \\
          & C2 &  0.5 $\pm$ 0.1 &  $-$              & $-$  \\
          & C3 &      $-$       &  3.4 $ \pm $  0.1 & $-$  \\
          & C4 &      $-$       &  0.4 $ \pm $  0.1 & $-$  \\
          & C5 &      $-$       &  0.4 $ \pm $  0.1 & $-$  \\
          & C6 &      $-$       &  0.4 $ \pm $  0.1 & $-$  \\
\hline
J1206+503 & C  & 3.4 $ \pm $  0.1 & 3.4  $\pm$  0.1  & 0.00  \\
\hline
J1313+507 & C  & 5.3 $ \pm $  0.1 &  3.5 $\pm$ 0.1  & 0.38  \\
          & C1 &      $-$         &  1.3 $\pm$ 0.1  & $-$   \\
\hline
J1315+516 & C  & 7.2  $\pm$ 0.1    &  6.2 $\pm$ 0.1 &  0.14  \\
          & C1 & 0.3  $\pm$ 0.1    &  $-$           &  $-$  \\
\hline
J1348+286& C & 17.7 $ \pm $ 0.4  & 24.4$\pm$ 0.1  & -0.29  \\
         & C1 & 2.4  $\pm$ 0.1   & 0.7 $\pm$ 0.1  &  1.12 \\
         & C2 & 1.3  $\pm$ 0.1   & 0.5 $\pm$ 0.1  & 0.87  \\
         & C3 & 0.7  $\pm$ 0.1   & $-$            & $-$   \\
         & C4 &      $-$         & 7.3 $\pm$ 0.1  & $-$   \\
         & C5 &      $-$         & 0.5 $\pm$ 0.1  & $-$   \\
         & C6 &      $-$         & 0.3 $\pm$ 0.1  & $-$   \\
         & C7 &      $-$         & 0.3 $\pm$ 0.1  & $-$   \\
\hline
\end{tabular}
\end{center} 
\label{table2}
\end{table}


\section{Results}
\subsection{Radio structures}
Here we present higher resolution VLBA observations of five
hybrid objects discovered by G06. Two of them revealed a core-jet structure
(J1154+513 \& J1348+286), and the other two show the probable presence of a weak jet (J1313+507
\& J1315+516). For J1206+503 we have only detected a point-like central component. 
We analyzed the flux densities of the central components of all five sources measured 
on VLA and VLBA images (Tables~1 \&
2). The summarized flux densities on VLBA scales of four hybrid sources
(J1206+503, J1313+507, J1315+516, \& J1348+286) are comparable to those
measured on VLA maps, although for J1348+286 the slightly resolved
structure suggests the presence of weak radio jet. The core flux density of
J1154+513 measured on
the VLA scale is higher than measured on VLBA images, and yet in this case the core has
been clearly resolved into a core-jet structure. We conclude that there
is no additional diffuse radio emission in the proximity of the nucleus of these
sources on $\sim1-10$\,kpc scale, which could be anticipated in the case of FR\,I radio galaxies. 
We suggest that  -
the potentially present - radio jets of the HYMORS  must be very weak on parsec
scales, similar to those observed in FR\,II radio galaxies.  

We have gathered the basic
properties of the new hybrid objects in Table 1. The spectroscopic or
photometric redshift is known for four out of five sources. Their estimated luminosity
exceeds the traditional FR\,I/FR\,II break luminosity, which at 178\,MHz is
$\sim 10^{25.5}~{\rm W~Hz^{-1}~sr^{-1}}$ \citep{fr}, this indicates that they have radio power
similar to FR\,IIs.
\subsection{Orientation}

In our analysis of the observed HYMORS, we focused on determining of  
the spatial orientation of the sources and compared the arcsecond-scale
resolution VLA images (G06) with the milliarcsecond-scale structures detected
during the VLBA observations. Since the radio morphology (the jet and
counter-jet visibility)  can give us a rough estimation of the orientation, we wanted
to check that the unusual hybrid morphology is connected with some specific
orientation towards the observer. Any clean-cut trend in the orientation could support a possible geometric
explanation of the HYMORS phenomenon similar to the one presented in \citet{mar12a,mar12b} for sources J1211+743,
3C\,249.1, and 3C\,334.
In the case of J1154+513 detected jet is clearly directed
towards an FR\,I-like part of the source (Fig.~\ref{hybrids_1}). 
In the case of J1348+286, the radio
structure on VLBA images suggests that the jet is pointing towards the FR\,II-like
side. Also in the case of suspected jets (J1313+507 \& J1315+516), none of large-scale
side is preferred.

The counterjet is not visible in any of the observed objects, which prevents us
from calculating the beaming from a known jet to counterjet brightness ratio.
However, to do this we estimated the Doppler factor using the method suggested by \cite{gio01}, which is
based on the correlation between the core and total radio power in radio
galaxies:
\begin{equation}
{\rm log{P_{ci5}}=0.62 log{P_{t}}+8.41},
\end{equation}
where $P_{ci5}$ is the intrinsic core 5\,GHz radio luminosity derived assuming
$\gamma=5$, and $P_{t}$ is the total radio luminosity at 408\,MHz. 
The Doppler factor is then estimated by using the $P_{ci5}$ and observed
5\,GHz radio core luminosity $P_{c,o}$ according to the equation;
\begin{equation}
{\rm P_{c,o}}={P_{ci5}\delta^{2+\alpha}}
\end{equation}
assuming spectral index $\alpha=0$. The observed    
5\,GHz radio core luminosity was calculated using VLA flux from G06 and with
a K-correction for $\alpha=0$. The 408\,MHz flux were taken from the
literature (J1348+286) or from an interpolation between 178/151\,MHz and
1.4\,GHz fluxes in other cases. The derived Doppler factor is less than 2.3
for all sources implying viewing angle more than $21^{\rm o}$ (for 
$\gamma$ = 1.5), which can easily prevent us from detecting the 
counter-jet on the parsec scale.


The above results indicate  low beaming in the 
five studied HYMORS and do not explain the mixture of FR\,I and FR\,II 
morphology. Moreover, the presented VLBA observations do not show
that there is any preferable spatial orientation in hybrid objects.

Candidate HYMORS in G06 were selected at at 1.4\,GHz,
a frequency at which Doppler boosting operates. The initial sample
constructed from FIRST survey consisted of more than 1700 sources selected in five subareas
($F_{1.4\,\mathrm{GHz}}\geq 20$\,mJy and angular size $\theta > 8^{''}$).
It is expected that the population of beamed objects is more numerous in the
sample constructed at  1.4\,GHz than in the case of the similar sample chosen at lower frequencies.



However, if we assume randomly oriented radio sources (e.g. \citet{hough92}, \citet{bar89}), 
then the number of objects with Doppler factor greater than 1.5 in an unbiased sample decreases from 
10\% to 8\%  (for $\gamma =$ 1.5 , 5 respectively). Therefore it is not surprising
that there is only one object in our sample with a Doppler factor greater than 2. The Doppler factor for 
quasar J1348+286 amounts to 2.2\,. 

 
\section{Discussion}

The radio observations of HYMORS show that the sources possess an FR\,I
radio morphology on one side of the core and an FR\,II type structure on the
other side (GKW00). In the case of FR\,II sources the jets are  weak but remain relativistic 
on all scales and  terminate abruptly at the end of the jet forming a hotspot . 
 A characteristic feature of FR\,I sources is the presence of prominent
twin jets that passing through a so-called flaring point inflate turbulent
lobes with plumes at the end and with no evidence of
the strong shocks and hotspots. The flaring point probably marks a transition
in the jet where the flow decollimates and starts to decelerate \citep{laing99,laing02}.
It also corresponds to a sudden increase in the rest-frame
emissivity. The length of the deceleration region depends on the total power of
the source, and ranges from $\sim$2\,kpc for luminosities $< 10^{24} {\rm
W~Hz^{-1}}$ at 1.4\,GHz to $\sim$10\,kpc for the stronger sources. 
The flaring points have also been observed
in WAT objects \citep{hasek04} and are described as hotspot-like features similar
to those visible in FR\,IIs.

\begin{figure*}
\centering
\includegraphics[width=\textwidth]{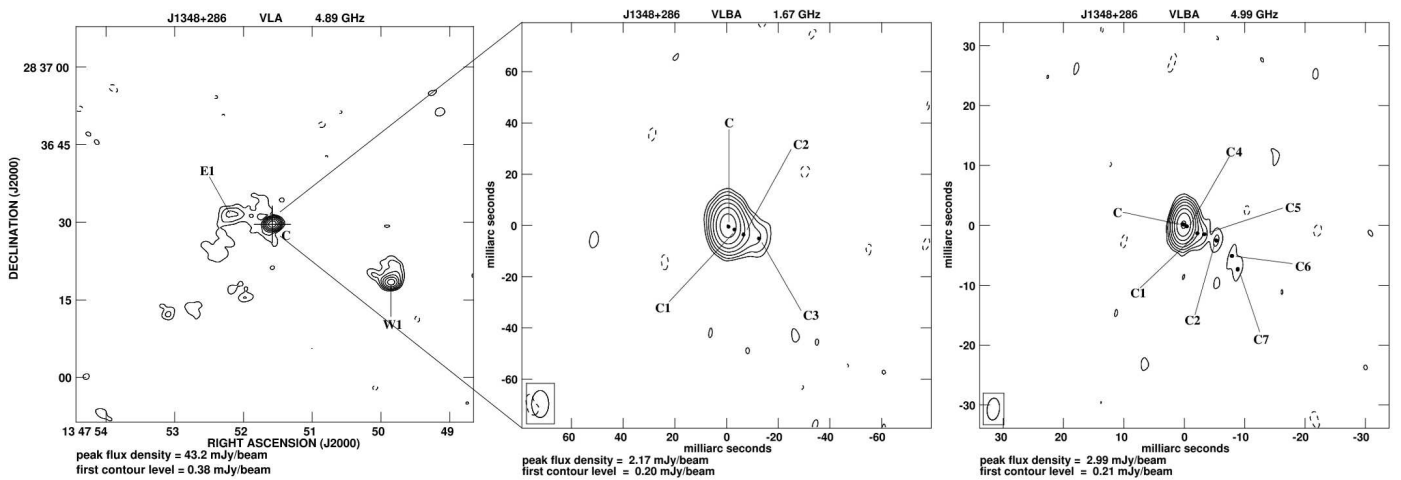} 
\caption{
VLA C-band images reprinted from G06 (left) and VLBA L-band and C-band images
(middle and right)of J1348+286. Contours increase  
by a factor 2, and the first contour level corresponds to $\approx  
3\sigma$. A cross indicates the position of an optical object found using
the most actual version of the SDSS.}   
\label{hybrids_3}
\end{figure*}

We find such hotspot-like components in the arcsecond-scale radio morphologies of HYMORS.
These compact, bright features are visible in the inner part of the
fuzzy FR\,I side in all cases (components W1 in the case of J1206+503 and
J1313+507, E1 in J1154+513, J1315+516, and
J1348+286, Figs.\,\ref{hybrids_1} - \ref{hybrids_3} in this paper). 
They lie closer to the radio core than the
hotspots from the FR\,II side and possess steep spectra.  
Both the FR\,I and FR\,II parts of the hybrid objects show bends and
disruption of the radio structure similar to those observed in the WAT
objects \citep{hasek04}. We speculate that the bright compact features
visible on the FR\,I side in our hybrid objects could be regions of
transition to FR\,I morphology, which originally started as FR\,IIs. Such a scenario is
considered in many numerical simulations that try to explain the development of the FR\,I
and FR\,II morphology. In the model of \cite{meliani} a powerful FR\,II-type
jet can be decelerated by a high density nonhomogeneous intergalactic
medium and
transform to FR\,I-type on one side and remains FR\,II on the other side. 
According to \cite{per12},  the FR\,II type jet can be disrupted by the growth
of helical instability. Indeed our high resolution VLBA observations of
HYMORS presented in this paper show similar parsec-scale weak jets on both sides of
the radio core.


We looked through the literature for more information about 
the milliarcsecond observations of hybrid objects. 
 The discussion about the HYMORS was started by GKW00 .
Since then several articles have appeared containing information about random discovery
of possible hybrid sources. Nevertheless, there are very few articles dedicated precisely to
HYMORS.
 Therefore, we focused our investigation on the primary sample of HYMORS defined by the
GKW00 and the one described in this paper. GKW00
made a search through the literature and classified six sources as HYMORS. Four out of six objects from
their sample have higher resolution VLBI
observations, and they possess core-jet structures with the jet
pointing towards the FR\,I side \citep{perez04,beasly02,giro04} or two-sided
jets \citep{lloyd02}. Two of the objects have been classified as
BL\,Lacs with small viewing angles \citep{wu07}.  
In the case of five
hybrid objects, classified by G06 and described there, large beaming
is excluded, and the detected parsec-scale jets suggest that neither the FR\,I-like nor the FR\,II-like 
side is preferred.   
Summing up, there is no complete morphological information about any of
the previous studied samples of HYMORS, and/or the classification as hybrid objects are
uncertain in some cases. That is why we draw conclusions about the
orientation and radio structures only of the HYMORS described in this
paper.
The new sample of hybrid objects presented here has, for the
first time, full information about their structure on different scales,
luminosity, and beaming estimations. 

Very recently the place of the HYMORS in the evolutionary scheme of AGNs has
been discussed by \citet{kunert_radio}. In this picture HYMORS are typical
FR\,II radio sources with disrupted jet structure on one side of the host
galaxy. The break-up of an evolutionary path between FR\,IIs and HYMORS should     
take place on scales $\sim10-15$\,kpc, during the compact steep spectrum   
(CSS) phase of AGN evolution. However, the reason for such behavior is not
obvious, and its explanation is connected with the  ongoing discussion
about the nature of the FR division.
The observations of the variety of the large-scale radio morphologies connected
with a wide range in AGN radio loudness suggest that the combination of multiple factors 
is needed to explain the origin of FR\,I/FR\,II dichotomy.
There is a growing number of publications suggesting that the luminosity threshold
($L_{178MHz}$) is either insufficient or misleading in terms of FR\,I/FR\,II dichotomy.
\cite{klc} show examples of sources that exhibit different morphological
classes than would be suggested of taking the luminosity criterion into account.
Moreover, X-ray and the optical spectroscopic studies \citep{best, butti} support the division of radio
sources based on line ratios (LEG and high excitation
galaxies - HEG types) rather then their observed morphology, and link the discussion to the
properties of central engine: black hole spin or mode of accretion. In this
classification, weak FR\,IIs (LEGs)  fueled via radiatively inefficient flows
at low accretion rates can be grouped together with FR\,Is
\citep[e.g.][]{rbs,best}. As has recently been shown, the HEG/LEG division
is present also among young radio objects: CSS and Gigahertz-Peaked Spectrum
(GPS) sources \citep{kunert_opto}, which are considered as progenitors of
large scale FR\,I and FR\,II sources.  
The above analysis and conclusions concerning the new sample of HYMORS
suggest they are FR\,IIs that, after the interactions with the inhomogeneous
medium, changed their morphology. However, the still open question is whether they
are FR\,II HEGs or LEGs.  Spectroscopic studies of HYMORS are
needed to determine the properties of their central engines, which could
appear an important element of the puzzle called HYMORS.

\section{Summary}
We have presented here the first high-resolution VLBA observations of five
hybrid sources that were discovered by us in order to determine if the unusual radio morphology is connected
with the orientation of objects towards the observer. 
The sample gives complete, multifrequency information about the radio structure 
of HYMORS on different scales, luminosity, and beaming estimations.
 These are  unbeamed objects with viewing angles more than {\bf $21^{\rm o}$} for the 
inner parsec scale structures.
Their VLBA observations do not suggest there is a preferable spatial orientation in hybrid objects. 
Two of them revealed milliarcsecond core-jet structure and another two probable weak jets.
We suggest that very weak parsec-scale radio jets could be present in all objects, which make 
HYMORS cores similar in properties to FR\,II radio galaxies. 
Moreover, the estimated luminosity of observed HYMORS
exceeds the traditional FR\,I/FR\,II break luminosity indicating they have
radio powers similar to FR\,IIs.

Based on the performed analysis of the sample of new hybrid objects we
suggested that HYMORS are FR\,IIs evolving in a heterogeneous environment.
However, this could be not the only factor making HYMORS different.  
The interesting question is wheter the hybrid objects are FR\,II HEGs or weaker
FR\,II LEGs that are grouped together with FR\,Is and seem to be on the
evolutionary path with CSS LEGs.


\begin{acknowledgements}
We thank an anonymous referee for carefully reading the manuscript and  
for the constructive comments.\\
The VLA and VLBA is operated by the U.S. National
Radio Astronomy Observatory, which is a facility of the National
Science Foundation operated under cooperative agreement by 
Associated Universities, Inc.
This research made use of the NASA/IPAC Extragalactic Database
(NED), which is operated by the Jet Propulsion Laboratory, California
Institute of Technology, under contract with the National Aeronautics
and Space Administration.
Use has been made of the fourth release of the Sloan Digital Sky
Survey (SDSS) Archive. Funding for the creation and distribution
of the SDSS Archive  was provided by the Alfred P. Sloan
Foundation, the Participating Institutions, the National Aeronautics
and Space Administration, the National Science Foundation, the U.S.
Department of Energy, the Japanese Monbukagakusho, and the Max
Planck Society. The SDSS Web site is http://www.sdss.org/.
The SDSS is managed by the Astrophysical Research Consortium
(ARC) for the Participating Institutions. The Participating Institutions
are The University of Chicago, Fermilab, the Institute for Advanced
Study, the Japan Participation Group, The Johns Hopkins University,
Los Alamos National Laboratory, the Max-Planck-Institute for
Astronomy (MPIA), the Max-Planck-Institute for Astrophysics
(MPA), New Mexico State University, University of Pittsburgh,
Princeton University, the United States Naval Observatory, and the
University of Washington.
\end{acknowledgements}

\end{document}